\documentclass[aps,preprint,showpacs,preprintnumbers,amsmath,amssymb,floatfix]{revtex4}

\usepackage{graphicx}
\usepackage{dcolumn}
\usepackage{bm}
\usepackage{longtable}

\begin{document}

\title{Spectral moment sum rules for the retarded Green's function and self-energy of the inhomogeneous Bose-Hubbard model in equilibrium and nonequilibrium}

\author{J.~K.~Freericks}\email{jkf@physics.georgetown.edu}
 \affiliation{Department of Physics, Georgetown University,
Washington, D.C. 20057}

\author{V.~Turkowski} \email{vturkows@mail.ucf.edu}
\affiliation{Physics Department and Nanoscience Technology Center, University of Central Florida, Orlando, Florida 32816}

\author{H.~R.~Krishnamurthy} \email{hrkrish@gmail.com}
\affiliation{Centre for Condensed Matter Theory, Department of Physics, Indian Institute of Science, Bangalore 560012, India}
\affiliation{Condensed Matter Theory Unit, Jawaharlal Nehru Centre for Advanced Scientific Research, Bangalore 560064, India}
\author{M. Knap}
\affiliation{Institute of Theoretical and Computational Physics, Graz University of Technology, 8010 Graz, Austria}

\date{\today}

\begin{abstract}
We derive expressions for the zeroth and the first three spectral moment sum rules for the retarded Green's function 
and for the zeroth and the first spectral moment sum rules for the retarded self-energy of the inhomogeneous 
Bose-Hubbard model in nonequilibrium, when the local on-site repulsion and  the chemical 
potential are time-dependent, and in the presence of an external time-dependent electromagnetic field.  We also
evaluate these moments for equilibrium, where all time dependence and external fields vanish. Unlike similar
sum rules for the Fermi-Hubbard model, in the Bose-Hubbard model case, the sum rules often depend on expectation values that cannot be determined simply from parameters in the Hamiltonian  like the interaction strength and chemical potential, but require knowledge of equal time many-body expectation values from some other source. We show how one can approximately evaluate these expectation values for the Mott-insulating phase in a systematic strong-coupling expansion in powers of the hopping divided by the interaction. We compare the exact moments to moments of spectral functions calculated from a variety of different approximations and use them to benchmark their accuracy.
\end{abstract}

\maketitle

\section{ Introduction}

The Bose-Hubbard model was originally introduced to describe the behavior of superfluids in disordered environments, where superfluidity could be lost by phase fluctuations induced via disorder (Bose glass)  and where potential-energy effects could localize the system into a Mott insulator~\cite{fisher_etal_1989}. While much work  ensued on examining the properties of this model within the context of disordered superconductors, it was later proposed that this model could describe the behavior of bosonic atoms placed on optical lattices~\cite{zoller_optical_lattice}, which was  experimentally realized soon thereafter~\cite{greiner}. Since the atoms in an optical lattice are also held to a small region of space via a trapping potential, the experimental systems are further complicated by being in an inhomogeneous environment.  This is sometimes viewed as an advantage, as one can approximately map out the homogeneous phase diagram from the inhomogeneous system if the local-density approximation holds, but it also complicates much of the theoretical analysis, because the systems are no longer translationally invariant, so momentum is no longer a good quantum number.

The model was extensively studied numerically even before the application to ultracold atoms was known. In this early work, the system was always assumed to be homogeneous.  Monte Carlo simulations in one dimension~\cite{scalettar} and two dimensions~\cite{trivedi}, density matrix renormalization group work~\cite{krish_dmrg,dmrg_monien}, as well as analytic approximations, like the strong-coupling approach~\cite{freericks_monien}, were carried out.  After the experimental measurements of the Mott insulator were completed~\cite{greiner}, much further work ensued, culminating in the current state-of-the-art in quantum Monte Carlo simulations~\cite{scalettar_rigol,trivedi2,prokofiev}, in density-matrix renormalization group calculations~\cite{dmrg_kollath} and in experimental determinations of the phase diagram, including corrections to the local density approximation~\cite{spielmann,chen,bloch,greiner2}.  While much is known about the phase diagrams, less is known about the many-body spectral functions and self-energies of the Bose-Hubbard model.  The many-body density of states (DOS) has been calculated via maximum entropy analytic continuation of quantum Monte Carlo (QMC) data~\cite{qmc_dos}, via the variational cluster approach (VCA)~\cite{knapp}, via the time-dependent density matrix renormalization group (DMRG) approach~\cite{gebhard} and approximately with a bosonic version of dynamical mean-field theory in the strong-coupling limit~\cite{vollhardt}.

Here, we describe how to derive exact sum rules for the spectral functions of the retarded Green's function (zeroth and first three moments) and of the retarded self-energy (zeroth and first moment).  We work with the most general case, which involves both spatially inhomogeneous and nonequilibrium (time-dependent) Hamiltonians, and also summarize the results for equilibrium and spatially homogeneous cases.  This work complements previous work we have done on the nonequilibrium spectral moment sum rules for the Fermi-Hubbard model~\cite{turkowski1,turkowski2,turkowski3}, which analyzed that model for the same general conditions.  But unlike the Fermi-Hubbard model case, where all of the sum rules for retarded functions could be evaluated simply from parameters in the Hamiltonian, in the case for the Bose-Hubbard model, the spectral moment sum rules relate to many-body averages that are nontrivial, and need to be determined from some independent calculation in order to fully evaluate the different sum rules.  This situation becomes more complicated for higher moments.  But, these expectation values involve equal time expectation values for the nonequilibrium situation, and static expectation values for the equilibrium case, so they are simpler and more accurate to calculate than the spectral functions themselves. For example, we illustrate how to approximately determine them for the Mott-insulating phase with a strong-coupling perturbation-theory expansion.

In Sec.~II, we introduce the model and describe the techniques used to solve for the spectral moment sum rules and we develop the sum rules for the moments of the Green's functions and self-energies. We also discuss the equilibrium homogeneous case, and present the moments as functions of momentum.  Numerical results comparing to different approximations for the spectral functions are performed in Sec.~III along with a strong-coupling expansion for the expectation values needed to determine the values of the moments.  Conclusions follow in Sec.~IV.

\section{ Formalism for the Bose-Hubbard model}
\label{Formalism}

The inhomogeneous nonequilibrium Bose-Hubbard model Hamiltonian can be written in the following general form in the Schr\"odinger representation:
\begin{eqnarray}
H(t)=-\sum_{ij}t_{ij}(t)b_{i}^{\dagger}b^{}_{j}
-\sum_{i}\mu_{i}(t)b_{i}^{\dagger}b^{}_{i}
+\frac{1}{2}\sum_{i}U_{i}(t)b_{i}^{\dagger}b^{}_{i}(b_{i}^{\dagger}b^{}_{i}-1),
\label{H}
\end{eqnarray}
where $\mu_{i}(t)$ and $U_{i}(t)$ can have arbitrary time dependence and can vary from site to site on the lattice. Similarly, the intersite hopping
matrix $-t_{ij}(t)$ can be time-dependent, such as in the case where there is an external electromagnetic field with arbitrary 
time-dependence described by a vector potential ${\bf A}({\bf r},t)$, and the bosons are charged, or in the case of cold atoms where the atoms move in a synthetic (Abelian) gauge field given by such a vector potential~\cite{spielman_vector}. In this case, we would have a specific form for the
time dependence of the hopping given by
\begin{eqnarray}
-t_{ij}(t)=-t_{ij}\exp \left( -i\int_{{\bf R}_{j}}^{{\bf R}_{i}}q{\bf A}({\bf r},t)d{\bf r}/\hbar c\right) ,
\label{At}
\end{eqnarray}
with $q$ the ``effective'' charge of the bosons and ${\bf R}_i$ the position vector for lattice site $i$.

The general form for the Hamiltonian given in Eq.~(\ref{H}) can be used to describe many different nonequilibrium and inhomogeneous situations.  For example, it can describe bosons on an optical lattice that have the optical lattice potential modulated at some frequency, and thereby modifying the hopping, chemical potential and interaction as functions of time (as is done, for example, in modulation spectroscopy~\cite{esslinger_modulation}); in this case, the local chemical potential would be described by the total chemical potential minus the local trapping potential at each lattice site.  It can describe bosons on an optical lattice that have the trap either oscillated or impulsed to create dipole oscillations.  It can also be used to describe a disordered Bose glass system (for a fixed distribution of local chemical potentials which represent the global chemical potential minus the diagonal on-site disorder).

Similar to the fermionic case, one can define a generalized nonlocal spectral function for the boson
retarded Green's function in the Heisenberg representation
\begin{equation}
G_{ij}^R(t_1,t_2)=G_{ij}^{R}(T,t) =-i\theta (t)\left\langle
\left[ b_{i}(T+t/2),b_{j}^{\dagger}(T-t/2)\right]\right\rangle \label{GR}
\end{equation}
in the following way:
\begin{eqnarray}
A_{ij}^{R}(T,\omega) =-
\frac{1}{\pi} {\rm Im}\int_{-\infty}^{\infty}dt e^{i\omega t}G_{ij}^{R}(T,t),
\label{AR}
\end{eqnarray}
where we have introduced Wigner's time variables: the average
time $T=(t_{1}+t_{2})/2$ and the relative time $t=t_{1}-t_{2}$ (with $t_1=T+t/2$ the time at which the boson destruction operator is evaluated and $t_2=T-t/2$ the time where the boson creation operator is evaluated). Note that we will often need to refer to the Green's function in both ways indicated in Eq.~(\ref{GR}), in terms of the times the operators are evaluated $t_1$ and $t_2$, or in terms of the average and relative times.  While we do not introduce a new notation for the reexpressed Green's function, the context for whether the time variables are the variables where the operators are evaluated or are the average and relative times is always obvious by the context of the equations and whether the time variables have a subscript or not. The square brackets are used to denote a commutator. The angular brackets denote a statistical average with respect to the initial equilibrium distribution of the system in the distant past, with a density matrix given by $\rho=\exp[-\beta H(t\rightarrow -\infty)]/Z$ and $Z={\rm Tr}\exp[-\beta H(t\rightarrow -\infty)]$.    Note that $T$ denotes the average time, not the temperature, which is denoted by $1/\beta$. The time dependence of the operators is with respect to the full Hamiltonian and is expressed in the Heisenberg representation via an evolution operator.

We further define the $n$th spectral moment of the corresponding spectral function in
Eq.~(\ref{AR}) to be
\begin{eqnarray}
\mu_{nij}^{R}(T)&=&\int_{-\infty}^{\infty}d\omega \omega^{n}
A_{ij}^{R}(T,\omega). \label{munR}
\end{eqnarray}
This spectral moment is connected to the bosonic Green's function in the following way:
\begin{eqnarray}
\mu_{nij}^{R}(T)= -\frac{1}{\pi}{\rm Im}
\int_{-\infty}^{\infty}d\omega  \int_{0}^{\infty}dt
e^{i\omega t} i^{n}\frac{\partial^{n}}{\partial t^{n}} G_{ij}^{R}(T,t) . \label{munR2}
\end{eqnarray}
From this equation, one can obtain the following formula:
\begin{equation}
\mu_{nij}^{R}(T)= -{\rm Im} \left[
i^{n}\frac{\partial^{n}}{\partial t^{n}} G_{ij}^{R}(T,t)
\right]_{t=0^{+}}.  \label{munR3}
\end{equation}
It is possible to show that the terms
proportional to the time derivatives of the theta function do not contribute to the moments.
Then, by using the Heisenberg equation of motion for the operators, one can obtain 
the following expressions for the zeroth
and the first three spectral moments of the Green's function:
\begin{eqnarray}
\mu_{0ij}^{R}(T)&=&\langle [b_{i}(T),b_{j}^{\dagger}(T)] \rangle , \label{mu0R} \\
\mu_{1ij}^{R}(T)&=&\frac{1}{2}\left[ \langle
[L^{1}b_{i}(T),b_{j}^{\dagger}(T)] \rangle -\langle
[b_{i}(T),L^{1}b_{j}^{\dagger}(T)] \rangle \right], \label{mu1R} \\
\mu_{2ij}^{R}(T)&=&\frac{1}{4}\left[ \langle
[L^{2}b_{i}(T),b_{j}^{\dagger}(T)] \rangle -2\langle
[L^{1}b_{i}(T),L^{1}b_{j}^{\dagger}(T)] \rangle +\langle
[b_{i}(T),L^{2}b_{j}^{\dagger}(T)] \rangle \right]
\nonumber \\
&~&+\frac{i}{4}\left[ \langle
[[b_{i}(T),H'(T)],b_{j}^{\dagger}(T)] \rangle +\langle
[b_{i}(T),[b_{j}^{\dagger}(T),H'(T)]] \rangle \right]
, \label{mu2R} \\
\mu_{3ij}^{R}(T)&=&\frac{1}{8}\left[ \langle
[L^{3}b_{i}(T),b_{j}^{\dagger}(T)] \rangle -3\langle
[L^{2}b_{i}(T),L^{1}b_{j}^{\dagger}(T)] \rangle +3\langle
[L^{1}b_{i}(T),L^{2}b_{j}^{\dagger}(T)] \rangle \right.
\nonumber \\
&~&\left. -\langle [b_{i}(T),L^{3}b_{j}^{\dagger}(T)] \rangle
\right]
\nonumber \\
&~&+\frac{i}{8}\left[ 
4\langle [[[b_{i}(T),H'(T)],H(T)],b_{j}^{\dagger}(T)] \rangle
-\langle [[[b_{i}(T),H(T)],H'(T)],b_{j}^{\dagger}(T)] \rangle
\right.
\nonumber \\
&~&\left. -3\langle [[[b_{i}(T),H'(T)],b_{j}^{\dagger}(T)],H(T)] \rangle
+3\langle [[[b_{i}(T),H(T)],b_{j}^{\dagger}(T)],H'(T)] \rangle
\right.
\nonumber \\
&~&\left.
-\langle [b_{i}(T),[[b_{j}^{\dagger}(T),H'(T)],H(T)] \rangle
-2\langle [b_{i}(T),[[b_{j}^{\dagger}(T),H(T)],H'(T)] \rangle
\right]
\nonumber \\
&~&-\frac{1}{8}\left[ 
\langle [[b_{i}(T),H''(T)],b_{j}^{\dagger}(T)] \rangle
-\langle [b_{i}(T),[b_{j}^{\dagger}(T),H''(T)]] \rangle
\right]
, \label{mu3R}
\end{eqnarray}
where
 $L^{n}O=[...[[O,H],H]...H]$ is the operation of commutation
performed $n$ times, and $H'(T)=dH(T)/dT$ and $H''(T)=d^{2}H(T)/dT^{2}$
are the first and the second derivative of the Hamiltonian in the Schr\"odinger picture with respect
to time which is then evaluated in the Heisenberg picture at the average time $T$ ({\it i. e.}, the derivative is taken before going to the Heisenberg picture). In equilibrium these derivatives are zero, but in the nonequilibrium
case  they can be finite due to the explicit time-dependence of the Hamiltonian parameters.
Evaluating the commutators is a straightforward, but tedious exercise.  Doing so results in the following expressions:
\begin{eqnarray}
\mu_{0ij}^{R}(T)&=&\delta_{ij},\label{mu0ij}
\\
\mu_{1ij}^{R}(T)&=&-\bar t_{ij}(T)+2U_{i}(T)n_{i}(T)\delta_{ij},\label{mu1ij} \\
\mu_{2ij}^{R}(T)&=&\sum_{l}\bar t_{il}(T)\bar t_{lj}(T)
-2U_{i}(T)n_{i}(T)\bar t_{ij}(T)-2\bar t_{ij}(T)U_{j}(T)n_{j}(T)
\nonumber \\
&~&-U_{i}^{2}(T)n_{i}(T)\delta_{ij} 
+3U_{i}^{2}(T)\langle n_{i}^2(T)\rangle
\delta_{ij},
\label{mu2ij}
\end{eqnarray}
\begin{eqnarray}
\mu_{3ij}^{R}(T)&=&-\sum_{l,m}\bar t_{il}\bar t_{lm}\bar t_{mj}(T)
\nonumber \\
&~&+2U_{i}n_{i}\sum_{l}\bar t_{il}\bar t_{lj}(T)+2\sum_{l}\bar t_{il}U_{l}n_{l}\bar t_{lj}(T)+2\sum_{l}\bar t_{il}\bar t_{lj}U_{j}n_{j}(T)
\nonumber \\
&~&+2U_{i}\delta_{ij}\sum_{l,n}\left( \bar t_{il}\bar t_{ln}\langle b_{i}^{\dagger}b_{n}\rangle 
                                     +\bar t_{li}\bar t_{nl}\langle b_{n}^{\dagger}b_{i}\rangle
                                     -2\bar t_{il}\bar t_{ni}\langle b_{n}^{\dagger}b_{l}\rangle\right)(T) 
\nonumber \\
&~&+U_{i}^{2}n_{i}\bar t_{ij}(T)+\bar t_{ij}U_{j}^{2}n_{j}(T)-3U_{i}^{2}\langle n_{i}^2\rangle \bar t_{ij}(T)-3\bar t_{ij}U_{j}^{2}\langle n_{j}^2\rangle (T)
\nonumber \\
&~&-4U_{i}\bar t_{ij}U_{j}\langle n_{i}n_{j}\rangle (T)-U_{j}\bar t_{ji}U_{i}\langle b_{j}^{\dagger}b_{i}b_{j}^{\dagger}b_{i}\rangle (T)
+\bar t_{ii}U_{i}^{2}n_{i}(T)\delta_{ij}
\nonumber \\
&~&+2U_{i}\delta_{ij}\sum_{l}U_{l}\left( \bar t_{il}\langle b_{i}^{\dagger}b_{l}\rangle
                                   +\bar t_{li}\langle b_{l}^{\dagger}b_{i}\rangle\right)(T)
-4U_{i}^{2}\delta_{ij}\sum_{l}\bar t_{li}\langle b_{l}^{\dagger}b_{i}\rangle (T)
\nonumber \\
&~&-2U_{i}\delta_{ij}\sum_{l}U_{l}\left( \bar t_{il}\langle b_{i}^{\dagger}b_{l}n_{l}\rangle
                                   +\bar t_{li}\langle n_{l}b_{l}^{\dagger}b_{i}\rangle
                                                                                   \right) (T)
+6U_{i}^{2}\delta_{ij}\sum_{l}\bar t_{li}\langle b_{l}^{\dagger}b_{i}n_{i}\rangle (T)
\nonumber \\
&~&-3\delta_{ij}U_{i}^{3}\langle n_{i}^2\rangle (T)
   +\delta_{ij}U_{i}^{3}n_{i}(T)+4\delta_{ij}U_{i}^{3}\langle n_{i}^3\rangle (T)
\nonumber \\
&~&+\frac{i}{4}\sum_{l}\left( \bar t_{il}'\bar t_{lj}-\bar t_{il}\bar t_{lj}'\right)(T)
-\frac{i}{2}[ (\bar t_{ij}'U_{j}-\bar t_{ij}U_{j}')n_{j}+(U_{i}'\bar t_{ij}-U_{i}\bar t_{ij}')n_{i}](T)
\nonumber \\
&~&+\frac{1}{4}\frac{d^{2}\bar t_{ij}(T)}{dT^{2}}-\frac{1}{2}\frac{d^{2}U_{i}(T)}{dT^{2}}n_{i}(T)\delta_{ij}
,
\label{mu3ij}
\end{eqnarray}
where $n_i(T)=\langle b_{i}^{\dagger}(T)b^{}_{i}(T)\rangle$, $-\bar t_{ij}(T)=-t_{ij}(T)-\delta_{ij}\mu_{i}(T)$, and the symbol $(T)$ reminds us that {\it both} the parameters in the Hamiltonian {\it and} the operator expectation values are to be evaluated at the average time $T$ in the expressions for the moments. Note that we have expressed the operator expectation values in the most compact form rather than subtracting off the average values to show the effects of correlations about the average values.  It is a simple, but tedious, exercise to convert to other forms if desired.

The expressions for the retarded
self-energy moments can be obtained by using the Dyson equation, which connects the retarded Green's function and self-energy
\begin{eqnarray}
G_{ij}^{R}(t_{1},t_{2})=G_{ij}^{R(0)}(t_{1},t_{2})+\sum_{l,m}\int dt_{3}\int dt_{4}G_{il}^{R(0)}(t_{1},t_{3})\Sigma_{lm}^{R}(t_{3},t_{4})G_{mj}^{R}(t_{4},t_{2}),
\label{Dyson}
\end{eqnarray}
where $G_{ij}^{R(0)}(t_{1},t_{2})$ is the noninteracting Green's function (in the case $U=0$), and $\Sigma^R_{ij}(t_1,t_2)$ is the retarded self-energy.
The first step is to rewrite 
Eq.~(\ref{Dyson})  in a combined frequency-average time
representation:
\begin{eqnarray}
G_{ij}^{R}(T,\omega )&=& G_{ij}^{R0}(T,\omega )+\sum_{lm}\int
d{\bar T}\int d{\bar t} \int
d\Omega \int d\nu e^{-i\Omega {\bar t}} e^{i\nu {\bar T}} \nonumber \\
&\times&G_{il}^{R0}\left( T+\frac{{\bar T}}{2}+\frac{{\bar
t}}{4},\omega +\Omega +\frac{\nu}{2}\right) \Sigma_{lm}^{R}
\left( T+{\bar T},\omega +2\Omega \right)G_{mj}^{R}\left(
T+\frac{{\bar T}}{2}-\frac{{\bar t}}{4},
\omega +\Omega -\frac{\nu}{2}\right). \nonumber \\
\label{Dysonnonequilibrium2}
\end{eqnarray}
where the internal time variables with the overbars should not be confused with the effective hopping matrix used to describe the Green's function moments.

One can represent the Green's function and the self-energy at large frequency in terms of an asymptotic $1/\omega$ expansion:
\begin{eqnarray}
G_{ij}^{R}(T,\omega )=\sum_{n=0}^{\infty}\frac{\mu_{nij}^{R}(T)}{\omega^{n+1}},
\label{Gmu}
\end{eqnarray}
\begin{eqnarray}
G_{ij}^{R(0)}(T,\omega )=\sum_{n=0}^{\infty}\frac{{\tilde \mu}_{nij}^{R}(T)}{\omega^{n+1}},
\label{G0mu}
\end{eqnarray}
\begin{eqnarray}
\Sigma_{ij}^{R}(T,\omega )=\sum_{n=0}^{\infty}\frac{C_{nij}^{R}(T)}{\omega^{n+1}}+\Sigma_{ij}(T,\omega =\infty).
\label{Sigmamu}
\end{eqnarray}
where ${\tilde \mu}_{ijn}^{R}(T)$ denotes the corresponding Green's function moments for the noninteracting Green's function (which has $U=0$) and $C_{nij}^{R}(T)$ are the moments of the retarded self-energy, defined via
\begin{equation}
C_{nij}^{R}(T)=-\frac{1}{\pi}{\rm Im}\int_{-\infty}^{\infty}d\omega \omega^n \Sigma^R_{ij}(T,\omega).
\end{equation}
The large-$\omega$ expansions in Eqs.~(\ref{Gmu})-(\ref{Sigmamu})
can be obtained by using the following spectral identities 
(valid for all retarded functions that decay rapidly enough for large relative time):
\begin{eqnarray}
G_{ij}^{R}(T,\omega )=-\frac{1}{\pi}\int_{-\infty}^{\infty}d\omega '
\frac{{\rm Im}G_{ij}^{R}(T,\omega ')}{\omega -\omega '+i0^+},
\label{Green function_A}
\end{eqnarray}
\begin{eqnarray}
\Sigma_{ij}^{R}(T,\omega )=-\frac{1}{\pi}\int_{-\infty}^{\infty}d\omega '
\frac{{\rm Im}\Sigma_{ij}^{R}(T,\omega ')}{\omega -\omega '+i0^+} +\Sigma_{ij}^R(T,\omega=\infty).
\label{Sigma_A}
\end{eqnarray}
At large $\omega$ the Green's function and self-energy must be real, or the spectral moments would diverge. The self-energy expansion in the last equation contains a frequency-independent term
$\Sigma_{ij}^{R}(T,\omega =\infty )$, which corresponds to the
constant (Hartree-Fock) term in the self-energy.

Then, one  inserts these expansions into
Eq.~(\ref{Dysonnonequilibrium2}) and considers separately all terms
which have the same order in $(1/\omega )$. Hence, it
is necessary to expand all the functions under the integrals in
powers of $(1/\omega )$. This leads the following equations which connect
the Green's functions and self-energy spectral moments:
\begin{eqnarray}
\mu_{0ij}^{R}={\tilde \mu}_{0ij}^{R}=\delta_{ij},
\label{Ceq1}
\end{eqnarray}
\begin{eqnarray}
\mu_{1ij}^{R}={\tilde \mu}_{1ij}^{R}+\sum_{l,m}{\tilde \mu}_{0il}^{R}\Sigma_{lm}^{R}(T,\omega =\infty )\mu_{0mj}^{R},
\label{Ceq2}
\end{eqnarray}
\begin{eqnarray}
\mu_{2ij}^{R}&=&{\tilde \mu}_{2ij}^{R}+\sum_{l,m}{\tilde \mu}_{0il}^{R}\Sigma_{lm}^{R}(T,\omega =\infty )\mu_{1mj}^{R}
+\sum_{l,m}{\tilde \mu}_{0il}^{R}C_{0lm}^{R}\mu_{0mj}^{R}\nonumber\\
&+&\sum_{l,m}{\tilde \mu}_{1il}^{R}\Sigma_{lm}^{R}(T,\omega =\infty )\mu_{0mj}^{R},
\label{Ceq3}
\end{eqnarray}
\begin{eqnarray}
\mu_{3ij}^{R}&=&{\tilde \mu}_{3ij}^{R}+\sum_{l,m}{\tilde \mu}_{0il}^{R}\Sigma_{lm}^{R}(T,\omega =\infty )\mu_{2mj}^{R}
+\sum_{l,m}{\tilde \mu}_{0il}^{R}C_{0lm}^{R}\mu_{1mj}^{R}+\sum_{l,m}{\tilde \mu}_{0il}^{R}C_{1lm}^{R}\mu_{0mj}^{R}
\nonumber \\
&+&\sum_{l,m}{\tilde \mu}_{1il}^{R}\Sigma_{lm}^{R}(T,\omega =\infty )\mu_{1mj}^{R}
+\sum_{l,m}{\tilde \mu}_{1il}^{R}C_{0lm}^{R}\mu_{0mj}^{R}+\sum_{l,m}{\tilde \mu}_{2il}^{R}\Sigma_{lm}^{R}(T,\omega =\infty )\mu_{0mj}^{R}.
\label{Ceq4}
\end{eqnarray}
We have suppressed the average time dependence of all of the spectral moments to save space.

From these equations, one can obtain the results for the retarded self-energy moments by using the results for the retarded Green's function
moments derived above for both finite and vanishing $U$:
\begin{eqnarray}
\Sigma_{ij}^{R}(T,\omega =\infty )=2U_{i}n_{i}\delta_{ij}(T),
\label{Sigma_infinity}
\end{eqnarray}
 \begin{eqnarray}
C_{0ij}^{R}(T)=-U_{i}^{2}n_{i}\delta_{ij}(T)-4U_{i}^{2}n_{i}^{2}\delta_{ij}(T)
+3U_{i}^{2}\delta_{ij}\langle n_{i}^2\rangle (T) ,
\label{C0}
\end{eqnarray}
\begin{eqnarray}
C_{1ij}^{R}&=&2U_{i}\delta_{ij}\sum_{l,n}\left( \bar t_{il}\bar t_{ln}\langle b_{i}^{\dagger}b_{n}\rangle 
                                     +\bar t_{li}\bar t_{nl}\langle b_{n}^{\dagger}b_{i}\rangle
                                     -2\bar t_{il}\bar t_{ni}\langle b_{n}^{\dagger}b_{l}\rangle\right)(T) 
\nonumber \\
&~&+4U_{i}\bar t_{ij}U_{j}(n_{i}n_{j}-\langle n_{i}n_{j}\rangle )(T)
\nonumber \\
&~&+\delta_{ij}U_{i}^{3}(4n_{i}^{2}+8n_{i}^{3}-12n_{i}\langle n_{i}^2\rangle 
   -3\langle n_{i}^2\rangle +n_{i} 
   +4\langle n_{i}^3\rangle )(T)
\nonumber \\
&~&-U_{j}\bar t_{ji}U_{i}\langle b_{j}^{\dagger}b_{i}b_{j}^{\dagger}b_{i}\rangle (T)
+\bar t_{ii}U_{i}^{2}n_{i}\delta_{ij}(T)
\nonumber \\
&~&+2U_{i}\delta_{ij}\sum_{l}U_{l}\left( \bar t_{il}\langle b_{i}^{\dagger}b_{l}\rangle
                                   +\bar t_{li}\langle b_{l}^{\dagger}b_{i}\rangle\right)(T)
-4U_{i}^{2}\delta_{ij}\sum_{l}\bar t_{li}\langle b_{l}^{\dagger}b_{i}\rangle (T)
\nonumber \\
&~&-2U_{i}\delta_{ij}\sum_{l}U_{l}\left( \bar t_{il}\langle b_{i}^{\dagger}b_{l}n_{l}\rangle
                                   +\bar t_{li}\langle n_{l}b_{l}^{\dagger}b_{i}\rangle
                                                                                   \right) (T)
+6U_{i}^{2}\delta_{ij}\sum_{l}\bar t_{li}\langle b_{l}^{\dagger}b_{i}n_{i}\rangle (T)
\nonumber \\ 
&~&-\frac{i}{2}[ (\bar t_{ij}'U_{j}-\bar t_{ij}U_{j}')n_{j}+(U_{i}'\bar t_{ij}-U_{i}\bar t_{ij}')n_{i}](T)
-\frac{1}{2}\frac{d^{2}U_{i}(T)}{dT^{2}}n_{i}(T)\delta_{ij},
\label{C1}
\end{eqnarray}
where, once again, the $(T)$ symbol is to remind us of the average time dependence of both the parameters of the Hamiltonian and of the operator expectation values.

The expressions for the spectral moment sum rules are complicated in general.  We want to summarize these results for the case of a homogeneous system in equilibrium, where there is no time dependence to the Hamiltonian, and where the hopping is the same value between all nearest neighbors, the chemical potential is uniform, as is the interaction energy $U$.
In this case, we can express the moments in momentum space, with respect to the time-translation-invariant momentum dependent Green's function
\begin{equation}
G_{\bf k}^R(t)=-i\theta(t)\langle [b_{\bf k}^{}(t),b^\dagger_{\bf k}(0)]\rangle,
\end{equation}
and self-energy
\begin{equation}
\Sigma_{\bf k}^R(\omega)=G^{R0}_{\bf k}(\omega)^{-1}-G^{R}_{\bf k}(\omega)^{-1},
\end{equation}
where the symbol $\omega$ is used for the Fourier transform from time to frequency space, which is used because the Hamiltonian is time independent in equilibrium.
The operator $b_{\bf k}$ is the momentum-space destruction operator, which satisfies $b_{\bf k}=\sum_i b_i \exp[-i{\bf k}\cdot {\bf R}_i]/\sqrt{N}$, with a corresponding formula for $b_{\bf k}^\dagger$. We find the Green's function moments become
\begin{eqnarray}
\mu_{0\bf k}^R&=&1,\\
\mu_{1\bf k}^R&=&\epsilon_{\bf k}-\mu+2Un,\\
\mu_{2\bf k}^R&=&(\epsilon_{\bf k}-\mu +2Un)^2+U^2\left ( 3\langle n^2\rangle-4n^2-n\right  ) ,\\
\mu_{3\bf k}^R&=&(\epsilon_{\bf k}-\mu+2Un)^3+U^2Zt_{ii+\delta} \langle n_ib_{i}^\dagger b_{i+\delta}^{}+b_{i+\delta}^\dagger b_{i}^{} n_i\rangle\nonumber\\
&+&U^2\epsilon_{\bf k}\left ( 6\langle n^2\rangle -12n^2-2n+4\langle n_in_{i+\delta}\rangle+\langle b_{i+\delta}^{\dagger 2}b_i^2\rangle\right )-3U^2\mu\left ( 3\langle n^2\rangle-4n^2-n\right )\nonumber\\
&+&U^3\left ( 4\langle n^3\rangle-8n^3 -3\langle n^2\rangle +n\right ) .
\end{eqnarray}
Here, the symbol $\delta$ denotes the nearest neighbor translation, $\langle n_in_{i+\delta}\rangle$ denotes the nearest-neighbor static density-density expectation value and $-t_{ii+\delta}$ is the nearest-neighbor hopping matrix element.  Since the system is homogeneous, the expectation values and the hopping matrix element are independent of {\it which} nearest neighbor translation $\delta$ is chosen. The high-frequency constant for the self-energy is $\Sigma(\omega\rightarrow\infty)=2Un$, and the self-energy moments become
\begin{eqnarray}
C_{0\bf k}^R&=&U^2 \left ( 3\langle n^2\rangle -4n^2- n\right ) ,\\
C_{1\bf k}^R&=&U^2Zt_{ii+\delta} \langle n_ib_{i}^\dagger b_{i+\delta}^{}+b_{i+\delta}^\dagger b_{i}^{} n_i\rangle+ U^2\epsilon_{\bf k}\left ( 4\langle n_in_{i+\delta}\rangle+\langle b_{i+\delta}^{\dagger 2}b_i^2\rangle -4n^2\right )\nonumber\\
&+&U^2\mu\left (-3\langle n^2\rangle+4n^2+n\right )+U^3\left (4\langle n^3\rangle+8n^3-3\langle n^2\rangle -12\langle n^2\rangle n+4n^2+n\right ).
\end{eqnarray}

The zeroth and first Green's function moments and the constant of the self-energy do not require any expectation values besides the filling.  The second Green's function and zeroth self-energy moments require just one correlation function
$\langle n^2\rangle -n^2$, while the higher moments require increasingly more (and more complex) correlation functions.

\section{Numerical Results}

There are not too many techniques which can accurately determine the Green's function and self-energy of the Bose Hubbard model for real frequencies.  To date, the main nonperturbative methods that have been tried include quantum Monte Carlo plus a maximum entropy analytic continuation~\cite{qmc_dos}, the variational cluster approach~\cite{knapp}, the time-dependent density matrix renormalization group~\cite{gebhard}, and a strong-coupling version of bosonic dynamical mean-field theory~\cite{vollhardt}. The dynamical mean-field theory calculation is most accurate in large dimensions, and we will not consider it further here. Both the QMC and VCA can be performed in any physical dimension, the density matrix renormalization group is limited to one dimension, which is where we will focus our attention first.  The quantum Monte Carlo approach can also calculate different static expectation values, so using this data will allow us to completely determine the different moments.

But, in general, it would be nice to have alternative methods to at least approximate the value of the spectral moment sum rules.  In the case of a Mott insulator with a small hopping, one can use strong-coupling perturbation theory in the hopping to calculate the different expectation values.  The Mott phase for a homogeneous system in equilibrium with vanishing hopping is given by
\begin{equation}
|n\rangle_0=\prod_{i=1}^N\frac{\left ( b^\dagger_i\right )^n}{\sqrt{n!}}|0\rangle,\label{eq: mott}
\end{equation}
where $|0\rangle$ is the vacuum state and the Mott phase has an average density of $n$. The energy of this state is $E_n^0=[-\mu n+Un(n-1)/2]N$. Since this state is nondegenerate, standard nondegenerate Rayleigh-Schr\"odinger perturbation theory can be used to find the wavefunction.  Proceeding in the canonical fashion, where the perturbed state $\langle n|$ satisfies $\langle n|n\rangle _0=1$, we find that the perturbed wavefunction satisfies
\begin{equation}
|n\rangle=|n\rangle_0+\frac{\hat Q_n}{E_n^0-\hat H_0}\hat V |n\rangle_0+
\frac{\hat Q_n}{E_n^0-\hat H_0}\hat V\frac{\hat Q_n}{E_n^0-\hat H_0}\hat V|n\rangle_0,\label{eq: pt_wf}
\end{equation}
through second order in the perturbation $\hat V$, since the first-order shift in the energy $_0\langle n|\hat V |n\rangle_0=0$ vanishes (the perturbation $\hat V$ is the hopping term in the Hamiltonian). Here $Q_n=\mathbb{I}-|n\rangle_0~_0\langle n|$ is the projector onto the space perpendicular to the unperturbed wavefunction $|n\rangle_0$. A straightforward calculation of the overlap of the perturbed wavefunction is
\begin{equation}
\langle n |n\rangle =1+\frac{Zt^2}{U^2}n(n+1)N+O\left ( \frac{t^4}{U^4}\right ).\label{eq: overlap}
\end{equation}
Here $Z$ is the number of nearest neighbors and $t$ is the magnitude of the nearest-neighbor hopping matrix element (we assume we are on a bipartite lattice so all odd order terms in the perturbation $\hat V$ vanish).

It is now a straightforward exercise to find the expectation values needed for the different moments.  We have
\begin{eqnarray}
\frac{\langle n|n_i|n \rangle}{\langle n|n\rangle}&=&n,\\
\frac{\langle n|n_i^2|n \rangle}{\langle n|n\rangle}&=&n^2+\frac{2Zt^2}{U^2}n(n+1)+O\left ( \frac{t^4}{U^4}\right ),\\
\frac{\langle n|n_i^3|n \rangle}{\langle n|n\rangle}&=&n^3+\frac{6Zt^2}{U^2}n^2(n+1)+O\left ( \frac{t^4}{U^4}\right ),\\
\frac{\langle n|n_in_{i+\delta}|n \rangle}{\langle n|n\rangle}&=&n^2-\frac{2t^2}{U^2}n(n+1)+O\left ( \frac{t^4}{U^4}\right ),\\
\frac{\langle n|n_i^{}b_i^\dagger b_{i+\delta}^{}|n\rangle}{\langle n|n\rangle}&=&\frac{2t}{U}n^2(n+1)+O\left ( \frac{t^3}{U^3}\right ),\\
\frac{\langle n|b_i^{\dagger 2} b_{i+\delta}^{2}|n\rangle}{\langle n|n\rangle}&=&\frac{t^2}{U^2}n(n+1)[n(n+1)+\frac{1}{2}(n-1)(n+2)]+O\left ( \frac{t^4}{U^4}\right ).
\end{eqnarray}
Here $\delta$ denotes a nearest neighbor translation, so that $i+\delta$ is a nearest neighbor site of site $i$ (the expectation values are independent of which neighbor because the problem we are solving is homogeneous).

We compare the accuracy of the above expectation values with ones calculated directly via a quantum Monte Carlo approach in Table~\ref{table: expectation_values} for a Mott insulator on a one-dimensional lattice with $t/U=0.05$, $n=1$, and low temperature.  One can see that the accuracy is excellent for the strong coupling perturbation theory in this parameter regime.

\begin{table}[t]
\caption{Comparison of expectation values calculated in strong coupling perturbation theory versus quantum Monte Carlo simulation.  The case considered is a one-dimensional Mott insulator with $n=1$, $t=0.05U$, and low temperature.\label{table: expectation_values}}
\begin{tabular} {l | c | c }
\hline
Expectation value & strong coupling & quantum Monte Carlo\\
\hline
$\langle n_i^2\rangle$ & 1.02 & $1.02\pm 0.0004$\\
$\langle n_i^3\rangle$ & 1.06 & $1.06\pm 0.001$\\
$\langle n_in_{i+\delta}\rangle$ & 0.99 & $0.9902\pm 0.0002$\\
$\langle n_i^{} b^\dagger_ib_{i+\delta}^{}\rangle$ & 0.20 &$0.2022\pm 0.0004$\\
$\langle b_i^{\dagger 2}b_{i+\delta}^2\rangle$ & 0.01 & not calculated\\
\hline
\end{tabular}
\end{table}

Using the strong-coupling perturbation theories, we find the momentum-dependent retarded Green's function moments approximately become
\begin{equation}
\mu_{2\bf k}^R=(\epsilon_{\bf k}-\mu+2Un)^2+U^2n(n+1)\left ( -1+\frac{6Zt^2}{U^2}\right ),
\label{eq: pt_mom2}
\end{equation}
\begin{eqnarray}
\mu_{3\bf k}^R&=&(\epsilon_{\bf k}-\mu+2Un)^3+4UZt^2n^2(n+1)-2U^2\epsilon_{\bf k}n(n+1)\nonumber\\
&\times&\left [ 1-\frac{2Zt^2}{U^2}-\frac{t^2}{U^2} [ n(n+1)-\frac{1}{2}(n-1)(n+2)]\right ]\nonumber\\
&+&3U^2\mu n(n+1)\left ( 1-\frac{6Zt^2}{U^2}\right )-U^3n(n+1)(4n-1)\left ( 1-\frac{6Zt^2}{U^2}\right ),
\label{eq: pt_mom3}
\end{eqnarray}
to order $t^4/U^4$.  The self-energy moments approximately become
\begin{equation}
C_{0\bf k}^R=-U^2n(n+1)\left ( 1-\frac{6Zt^2}{U^2}\right ),
\label{eq: pt_semom0}
\end{equation}
and
\begin{eqnarray}
C_{1\bf k}^R&=&4UZt^2n^2(n+1)-t^2\epsilon_{\bf k}n(n+1)[8-n(n+1)-\frac{1}{2}(n-1)(n+2)]\nonumber\\
&+&U^2\mu n(n+1)\left ( 1-\frac{6Zt^2}{U^2}\right )+U^3n(n+1)\left ( 1-\frac{6Zt^2}{U^2}\right ),
\label{eq: pt_semom1}
\end{eqnarray}
also to order $t^4/U^4$.

\begin{figure}[htb]
\vskip -1.0in
\includegraphics[scale=0.45]{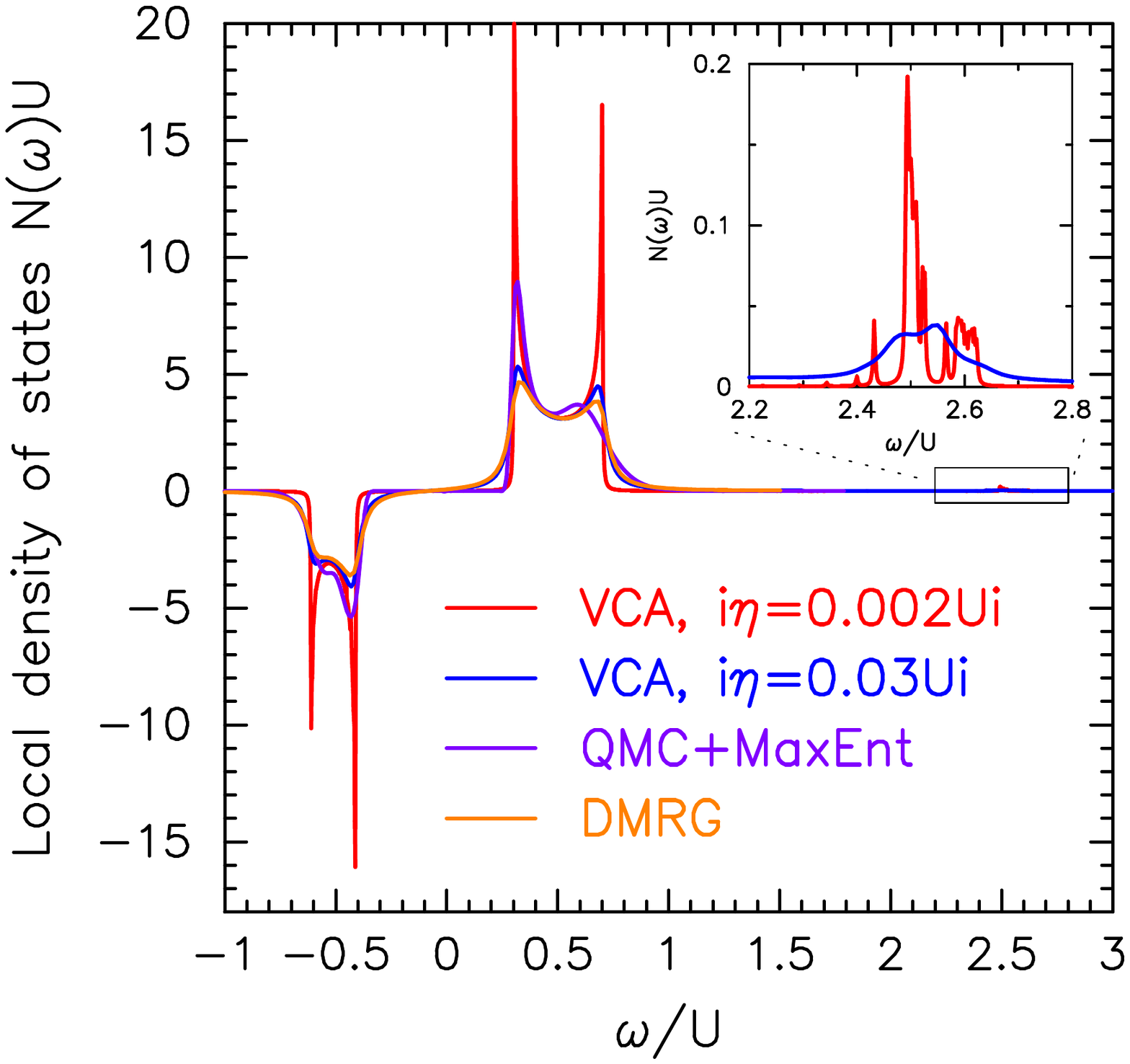}
\caption{(Color online.) Local density of states for the one-dimensional Bose Hubbard model in the Mott-insulating phase
with $n=1$.  The parameters are $t/U=0.05$ and $\mu/U=0.5$.  The energies are measured in units of $U$. We compare the variational cluster approach with two different broadenings (red and blue) to the quantum Monte Carlo plus maximum entropy approach (purple)  and to the density matrix renormalization group approach (orange). The inset zooms in on the region just above $2U$, where the variational cluster approach has some structure which is needed to get high
precision to the different moment sum rules. The data for the other two methods is cutoff before this frequency.}
\label{fig: dos}
\end{figure}

Now we are ready to compare the accuracy of different exact methods in calculating the many-particle density of states for the Bose-Hubbard model.  Our first test case is the Mott insulating phase in the one-dimensional model with $t/U=0.05$, $n=1$, and $\mu/U=0.5$ In Fig.~\ref{fig: dos}, we plot the local density of states for the three different methods that have been used for this problem: (i) the VCA at zero temperature with a Lorentzian broadening of $\eta=0.03$ and $\eta=0.002$~\cite{knapp}; (ii) QMC simulation plus maximum entropy analytic continuation~\cite{qmc_dos}, where the calculations are performed at a temperature $\beta=192$; and (iii) time-dependent density matrix renormalization group calculations~\cite{gebhard} at zero temperature with open boundary conditions and a Lorentzian broadening of $\eta=0.04$. One can see that there is a significant discrepancy between these different curves (in a pointwise fashion) but as we will see the moments are all quite close to one another.  This tells us that the density of states is quite sensitive to the broadening chosen, and it is difficult to tell which of these different techniques is most accurate (although, the quantum Monte Carlo technique uses the most unbiased algorithm to determine the density of states).  It is also apparent that one properly sees the correct gap in the density of states only with the methods that use the least broadening, as expected.

\begin{figure}[htb]
\vskip -1.0in
\begin{minipage}[h]{0.99\linewidth}
\centerline{\includegraphics[scale=0.34]{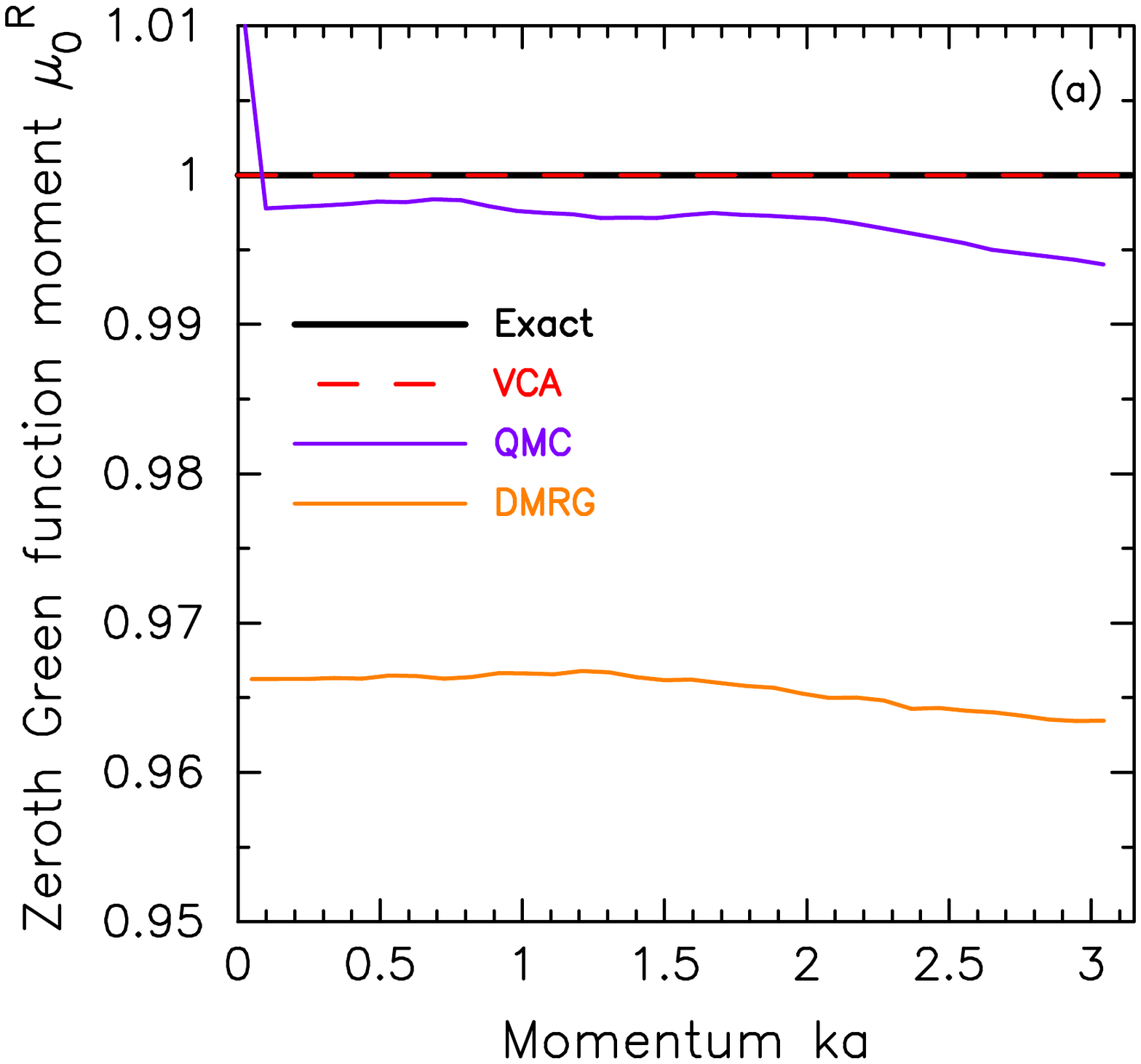}
\includegraphics[scale=0.34]{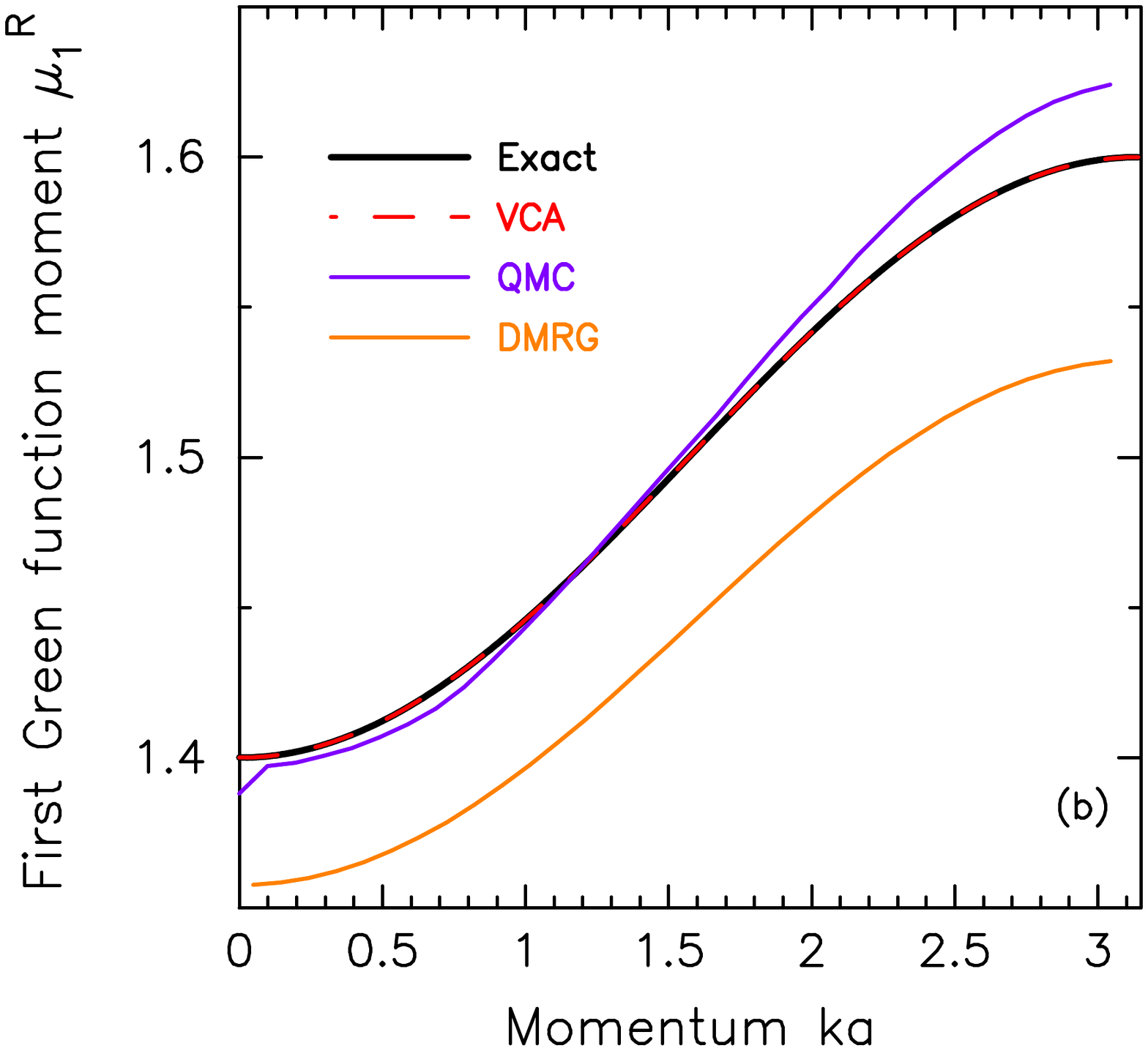}}
\end{minipage}
\vskip -1.5in
\begin{minipage}[h]{0.99\linewidth}
\centerline{\includegraphics[scale=0.34]{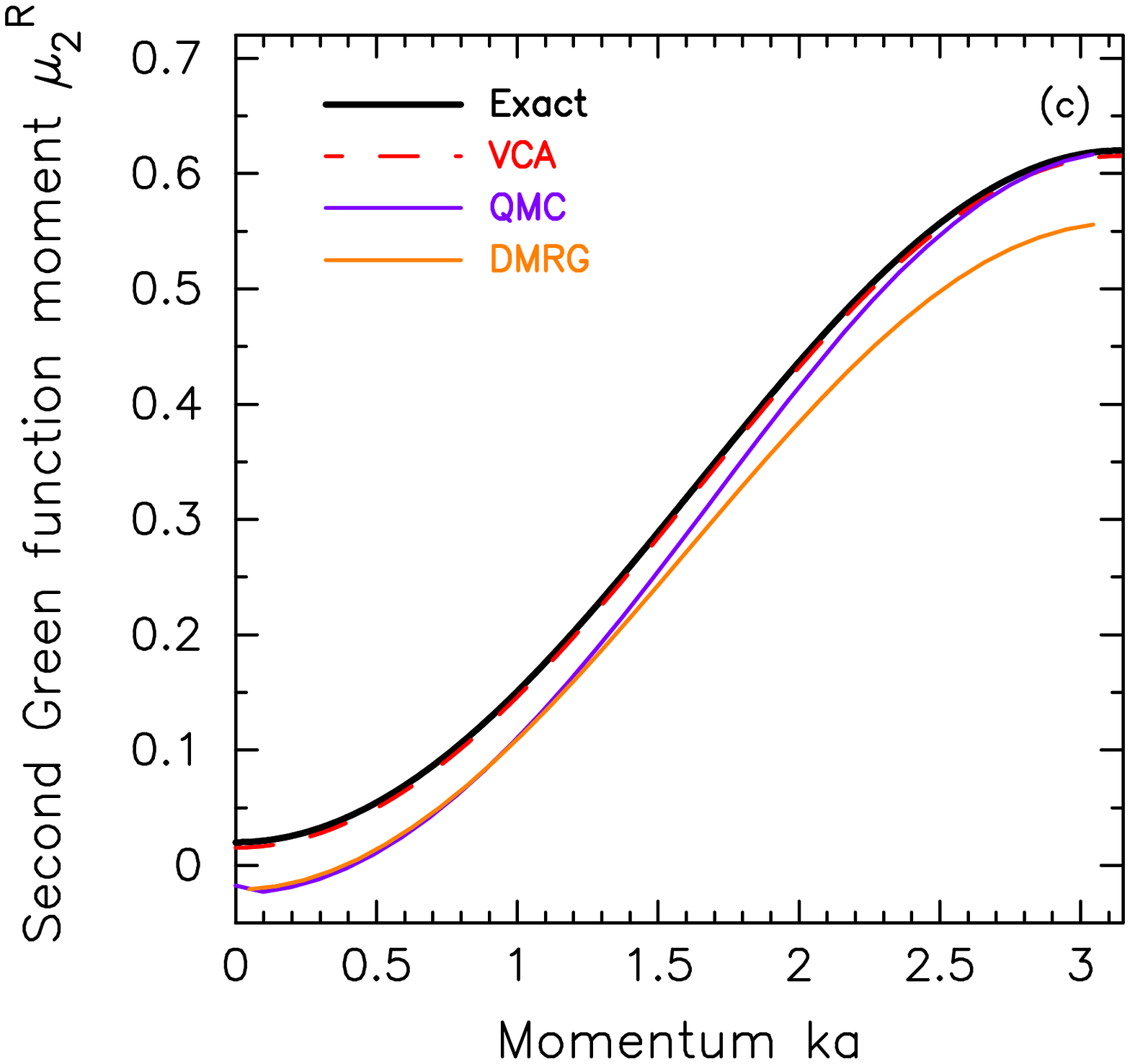}
\includegraphics[scale=0.34]{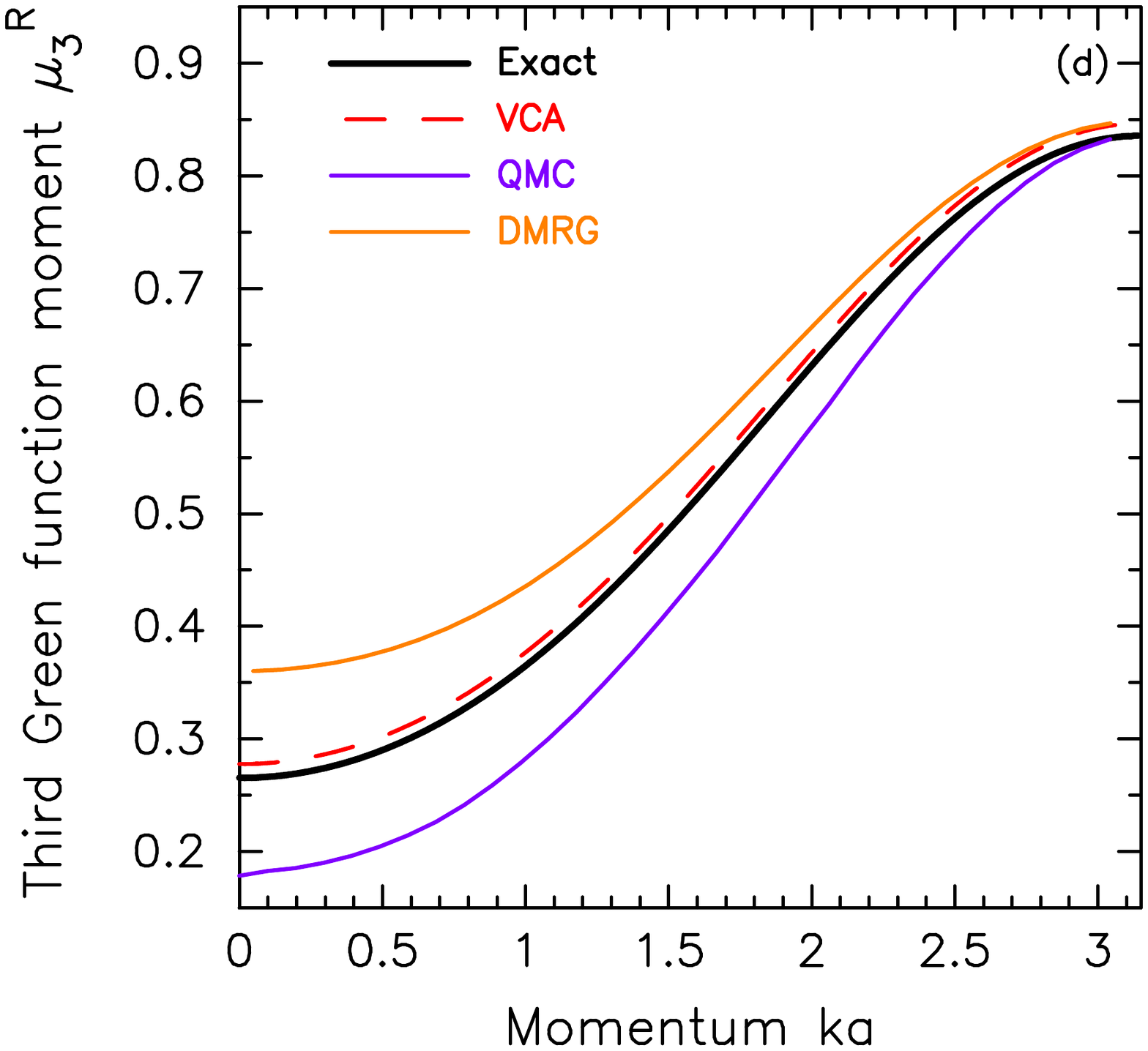}}
\end{minipage}
\caption{(Color online.) Spectral moments of the retarded Green's function as a function of momentum for the one-dimensional Bose Hubbard model in the Mott-insulating phase
with $n=1$.  The parameters are $t/U=0.05$ and $\mu/U=0.5$.  We compare the exact result (black) to the variational cluster approach  (red dashed line), to the quantum Monte Carlo plus maximum entropy approach (purple),  and to the density matrix renormalization group approach (orange). Panel (a) is the zeroth, panel (b) the first, panel (c) the second and panel (d) the third moment. The accuracy of the VCA is so good, one cannot see any deviation from the exact result in panels (a) and (b).  The quantum Monte Carlo and density matrix renormalization group results have higher errors.}
\label{fig: moment_gf}
\end{figure}

Now we move on to the task of comparing the different spectral moment sum rules.  We begin with the zeroth moment sum rule of the retarded Green's function, which essentially tests whether the system has conserved the correct number of states in the given calculation.  For the variational cluster approach, since it determines the Green's function as a set of delta functions and weights, we perform the integration of the moments via summing the relevant weights of the delta functions rather than introducing any artificial broadening into the calculation.  Doing this, appears to greatly improve the accuracy of the spectral moments themselves.  In Fig.~\ref{fig: moment_gf}~(a) we plot the exact result against the three different approximation techniques.  On the graph, one cannot see any error between the VCA and the exact result.  The quantum Monte Carlo is about a half a percent error, while the density matrix renormalization group results are about three and a half percent error.

The first retarded Green's function moment is plotted in Fig.~\ref{fig: moment_gf}~(b). Here the VCA and the QMC have small errors, with the former being less than 0.1\% and the latter on the order of a few percent.  The density matrix renormalization group result has the right shape, but appears to be systematically shifted off of the correct result causing about a 7\% error.

The second retarded Green's function moment is plotted in Fig.~\ref{fig: moment_gf}~(c).  This is the first moment that depends on a correlation function.  Once again, the VCA has superior accuracy, and the density matrix renormalization group results look like they are systematically shifted from the correct answer (so much so that at small momentum they have the wrong sign for the moment). 

Finally, in Fig.~\ref{fig: moment_gf}~(d), we plot the third moment sum rules. Here the deviations of all of the approximations are larger, but the VCA is clearly superior.  One might ask why the VCA appears to be so much better than the other two techniques, at least when we compare the moment sum rules?  We believe the answer to this lies in the inset in Fig.~\ref{fig: dos}.  There one can see that the VCA has some spectral weight at high energy, corresponding to higher on-site occupation of bosons.  The QMC and density matrix renormalization group results are cut off at lower frequencies, so they do not have this extra feature. This feature becomes increasingly important for higher moments, since the integrands are weighted more and also for even moments, since it can modify the cancellation that occurs between the positive and negative branches of the density of states.  In this sense, the moment sum rules can be a very delicate test of the accuracy of the different numerical calculations. In addition, the VCA, being based on a strong-coupling approach, is more accurate for large $U$.  We would expect other techniques to become more accurate as we move closer to the critical point and beyond.

We also want to test the spectral moment sum rules of the self-energy.  Here we have to now further process the data, as one cannot compute the self-energy solely from the density of states or the momentum-dependent spectral functions.  Instead, we use a Kramers-Kronig relation on the spectral functions to determine the momentum-dependent retarded Green's functions (the imaginary part is just $-\pi$ times the spectral function). Then we use Dyson's equation to extract the self-energy.  As a test, we compare the constant term of the real part of the self-energy to its exact result.  For the quantum Monte Carlo and for the density matrix renormalization group data, the extent of our data is too small in frequency for us to properly reach the limit where we can accurately extract the constant, but the error in the constant is less than 15\%.  Once we have the imaginary part of the self-energy, we simply integrate it times the appropriate power of frequency to see the sum rule.  For the VCA, we can no longer do this with the delta function representation, so we instead use the smaller broadening $\eta=0.002U$ data, which then leads to some noisy fluctuations in the integrated self-energy moments.  But the total noise level is not too high.

\begin{figure}[htb]
\vskip -1.0in
\begin{minipage}[h]{0.99\linewidth}
\centerline{\includegraphics[scale=0.34]{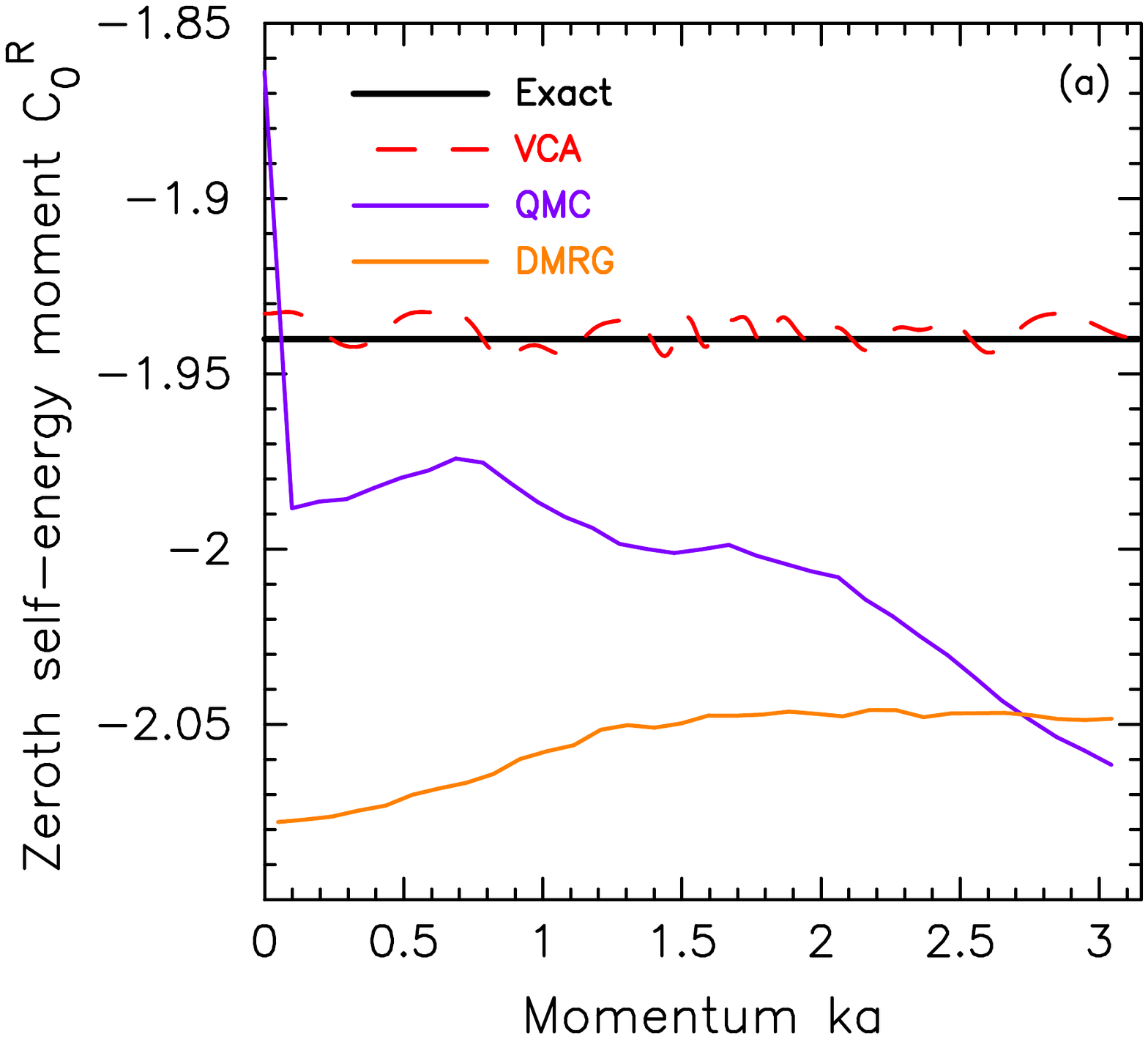}
\includegraphics[scale=0.34]{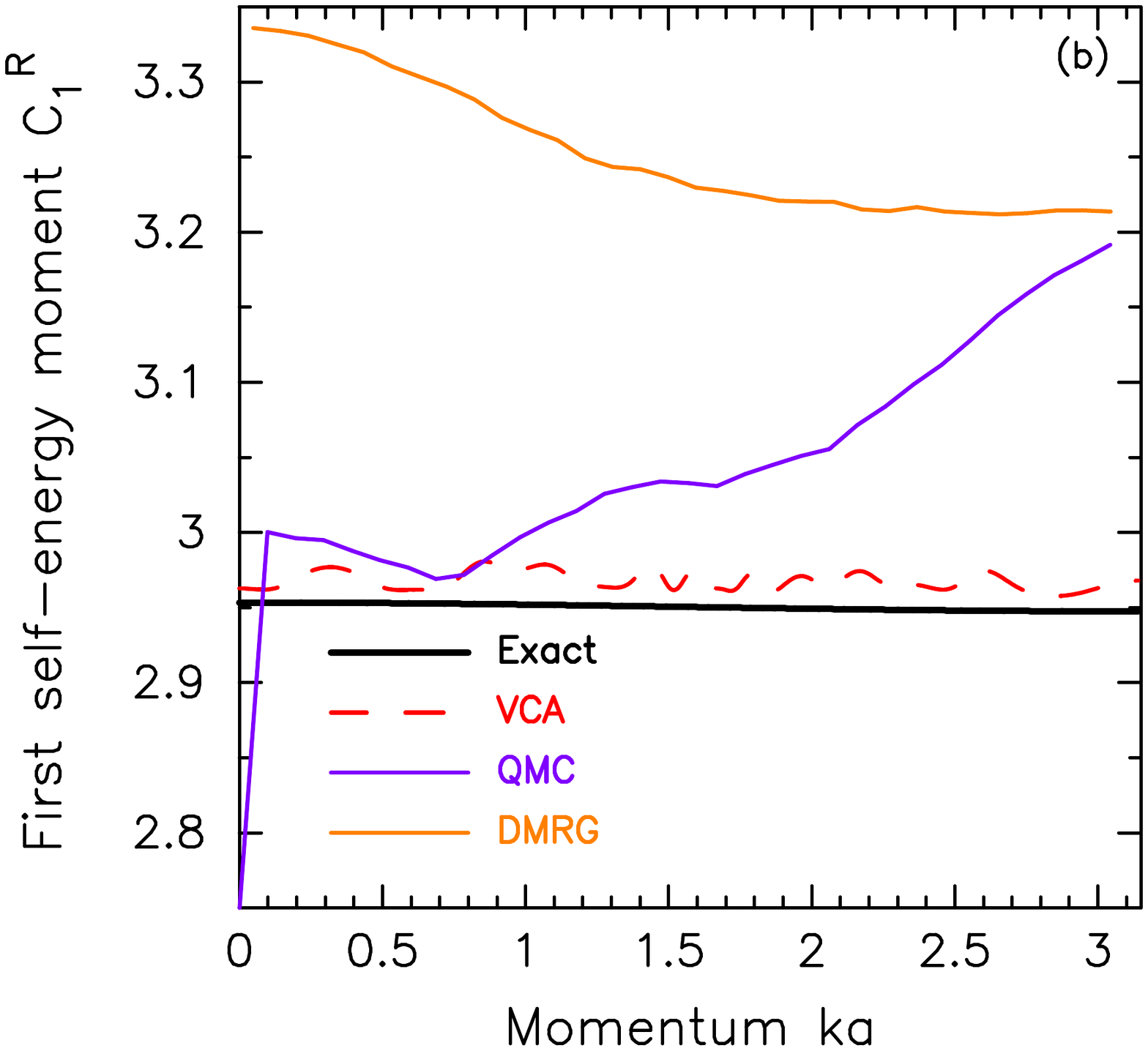}}
\end{minipage}
\caption{(Color online.) Zeroth (a) and first (b)  spectral moments of the retarded self-energy as a function of momentum for the one-dimensional Bose Hubbard model in the Mott-insulating phase
with $n=1$.  The parameters are $t/U=0.05$ and $\mu/U=0.5$, with energies measured in units of $U$.  We compare the exact result (black) to the variational cluster approach  (red dashed line), to the quantum Monte Carlo plus maximum entropy approach (purple),  and to the density matrix renormalization group approach (orange). Deviations are visible for all approximations.}
\label{fig: moment_se}
\end{figure}

We plot the zeroth spectral moment sum rule for the retarded self-energy as a function of momentum in Fig.~\ref{fig: moment_se}~(a). The sum rule itself is independent of momentum, even though the self-energies do vary with momentum.  The VCA has errors at the 0.5\% level.  The density matrix renormalization group has errors about ten times larger, and once again, there appears to be a systematic error in that data pushing it to slightly lower values.

Finally, we plot our last one-dimensional result, the first moment spectral sum rule of the retarded self-energy versus momentum in Fig.~\ref{fig: moment_se}~(b). Here we see similar results as with the zeroth moment, perhaps with somewhat larger errors.

The conclusion of this work on the one-dimensional example is that the spectral moment sum rules for the retarded Green's functions and for their self-energies provides useful data to help us predict the accuracy of different numerical calculations.  While they cannot provide us with sufficient data to determine what the appropriate widths of different spectral features should be in the Green's functions by examining pointwise values of the spectral functions, they do tell us important information about the weight under the curves and of their respective shapes.  Indeed, the higher moments, are particularly sensitive to small weight structures at high energy and could help identify whether approximations are cut off at too small a frequency and missed some higher energy spectral weight.

\begin{figure}[htb]
\vskip -1.0in
\begin{minipage}[h]{0.99\linewidth}
\centerline{\includegraphics[scale=0.34]{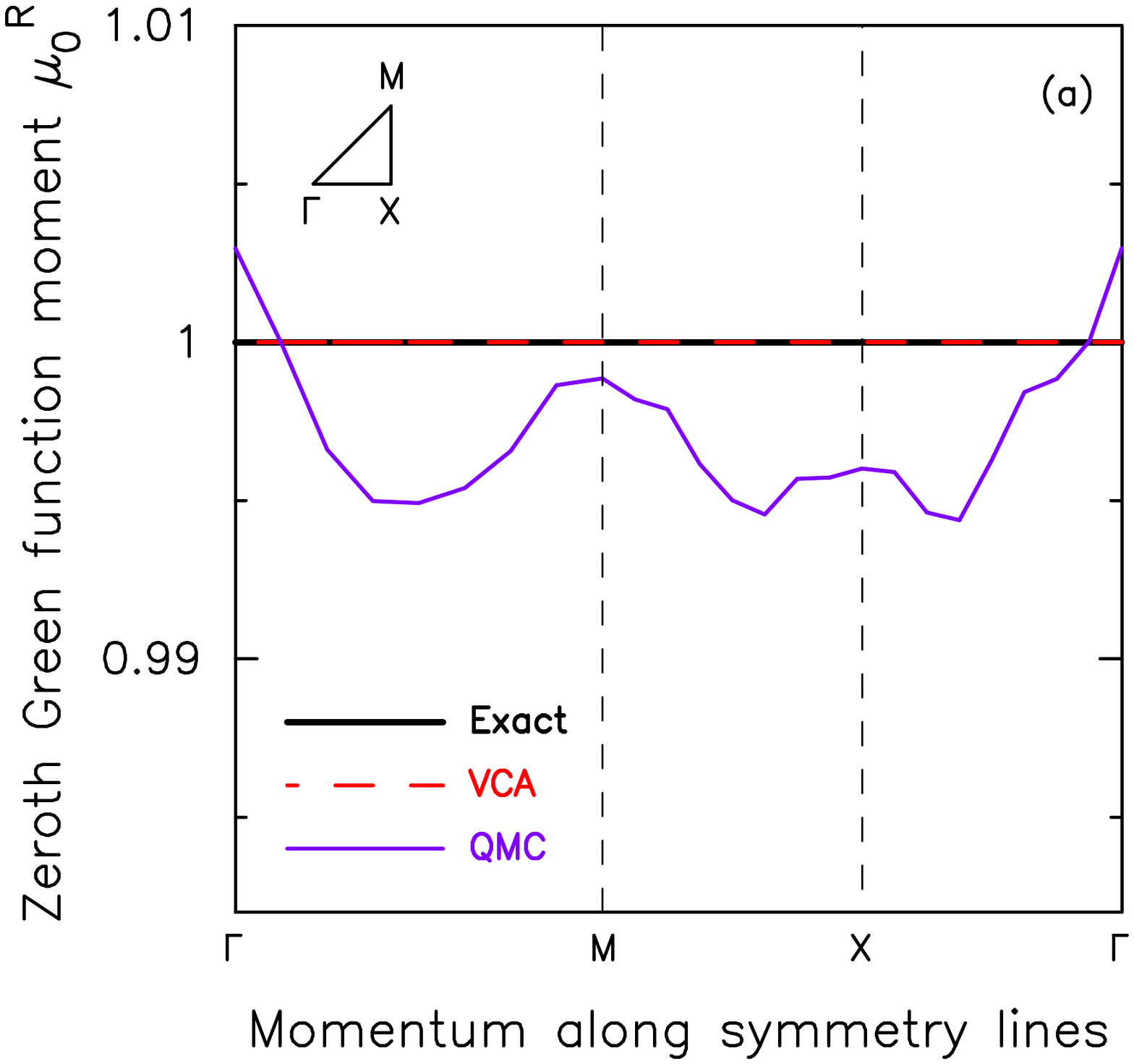}
\includegraphics[scale=0.34]{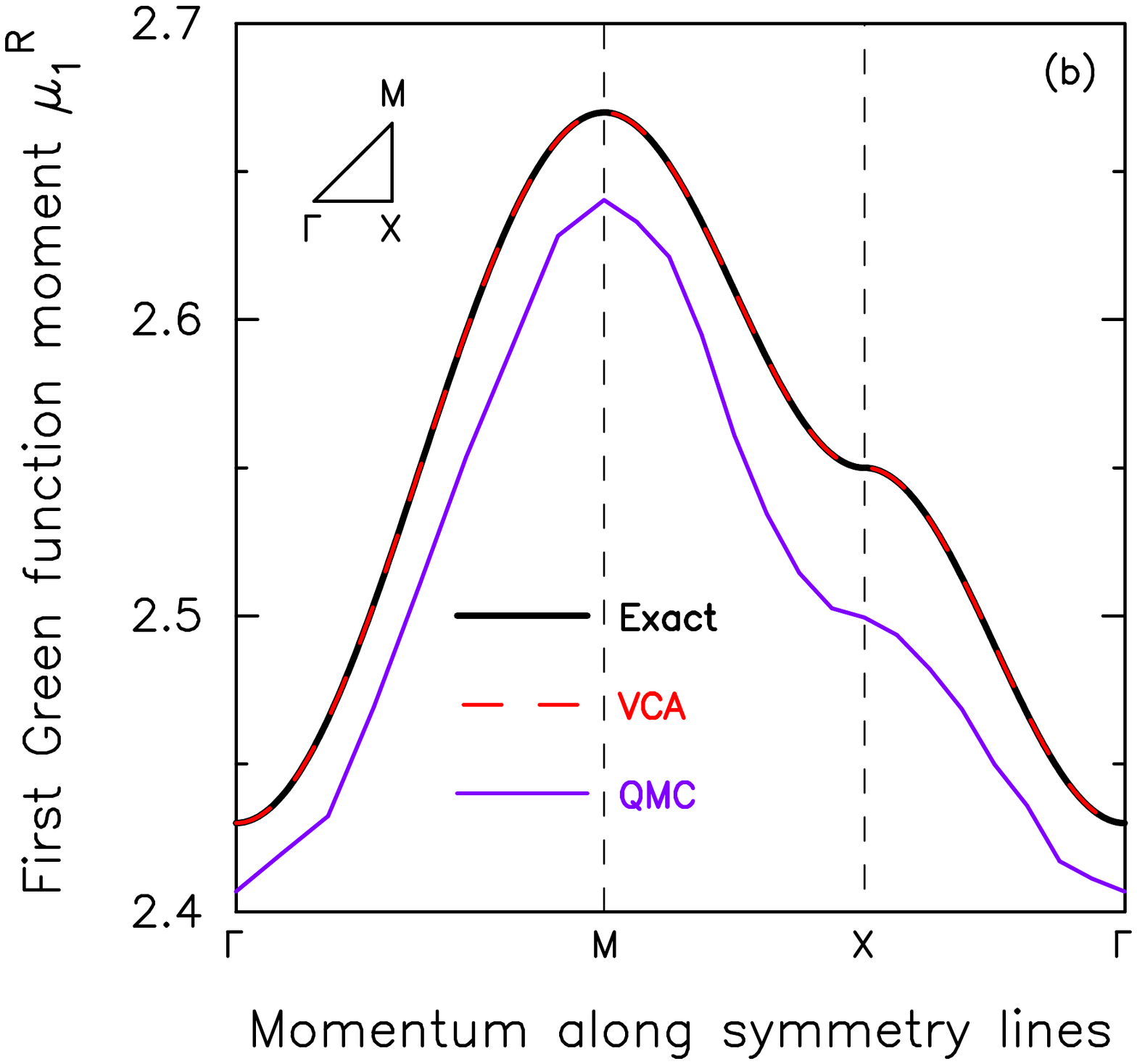}}
\end{minipage}
\vskip -1.5in
\begin{minipage}[h]{0.99\linewidth}
\centerline{\includegraphics[scale=0.34]{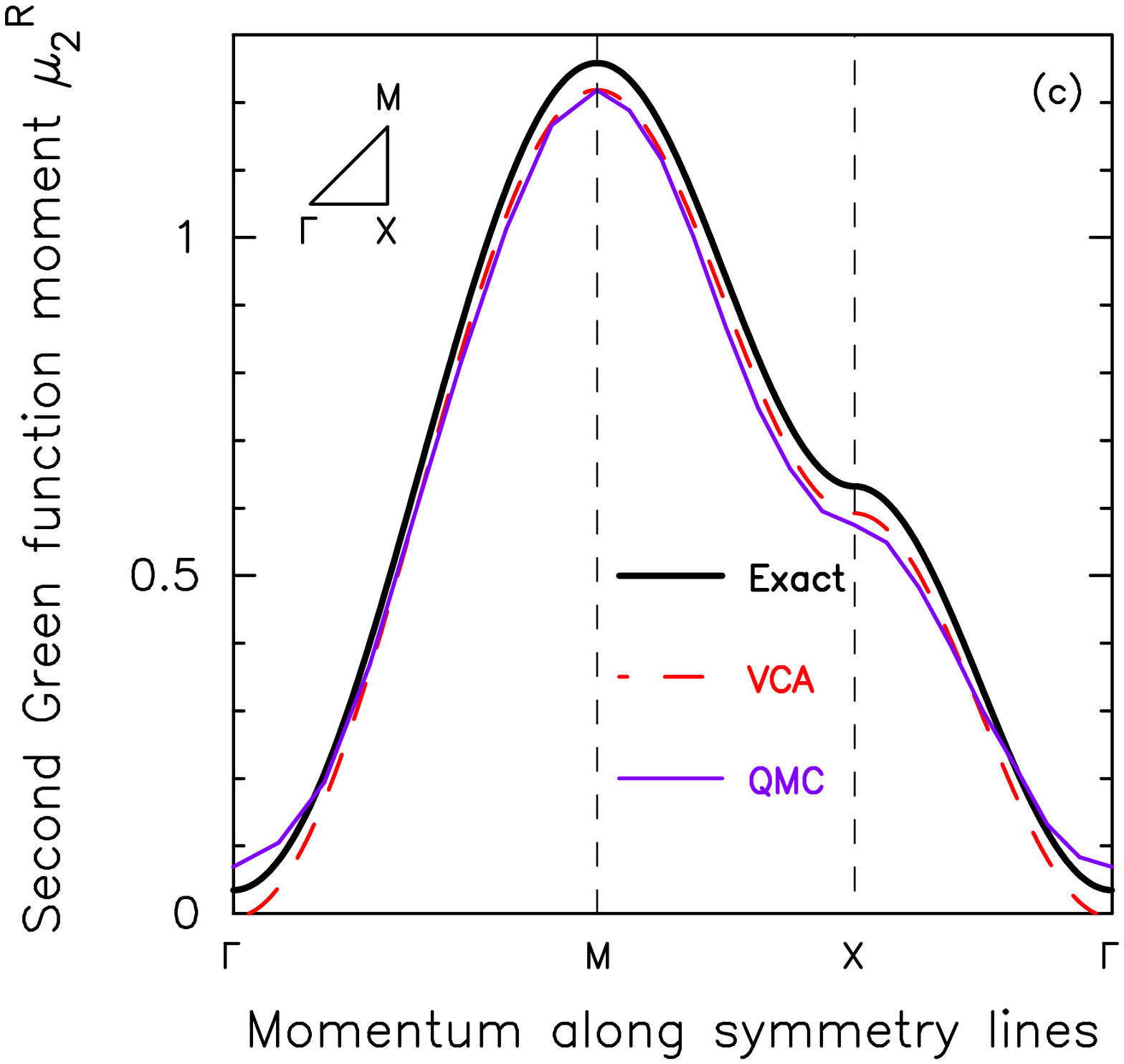}
\includegraphics[scale=0.34]{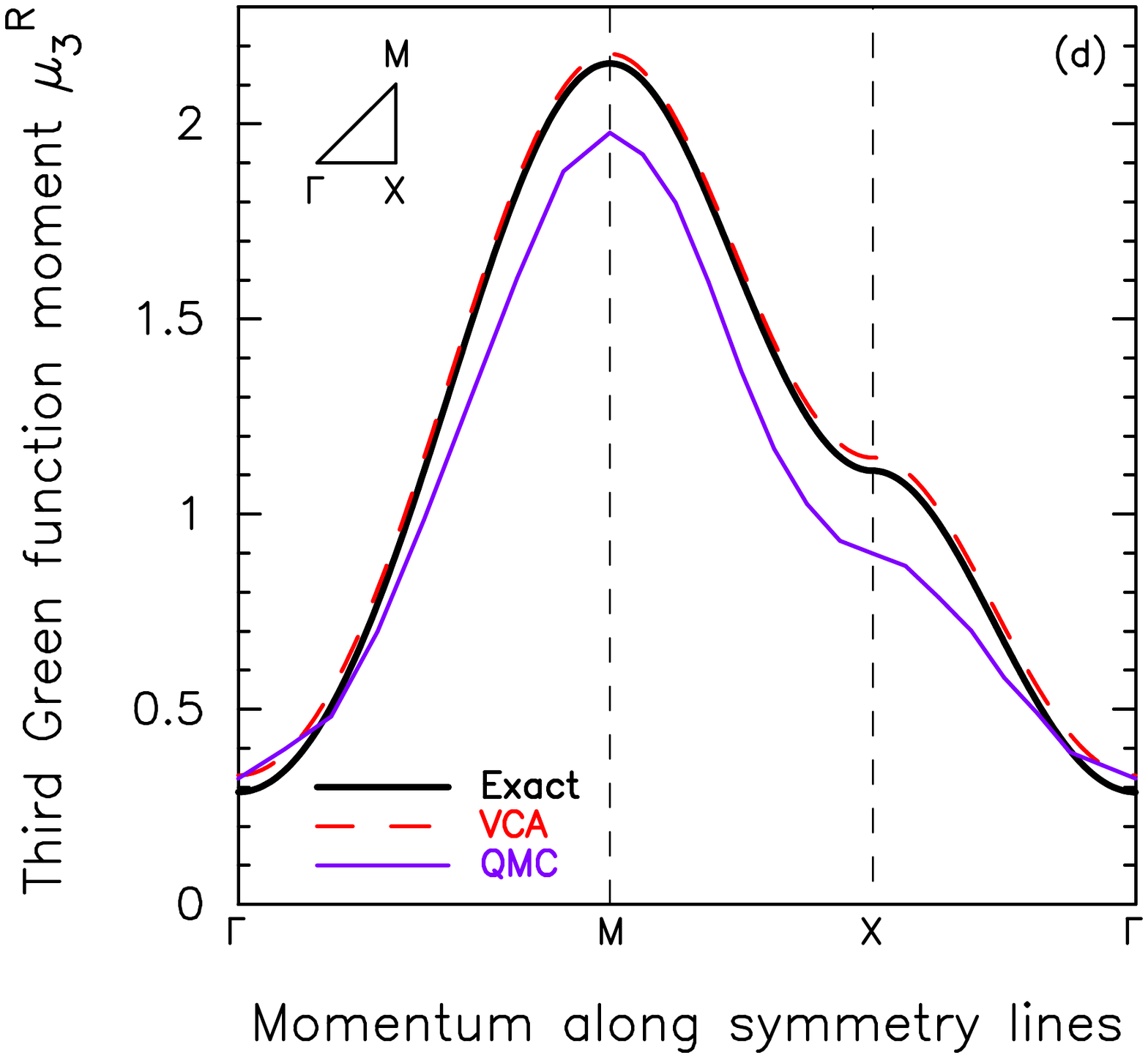}}
\end{minipage}
\caption{(Color online.) Spectral moments of the retarded Green's function as a function of momentum for the two-dimensional Bose Hubbard model on a square lattice in the Mott-insulating phase
with $n=2$.  The parameters are $t/U=0.03$ and $\mu/U=1.45$.  We compare the exact result (black) to the variational cluster approach  (red dashed line), and to the quantum Monte Carlo plus maximum entropy approach (purple). Panel (a) is the zeroth, panel (b) the first, panel (c) the second and panel (d) the third moment. The accuracy of the VCA is so good, one cannot see any deviation from the exact result in panels (a) and (b).  The quantum Monte Carlo results have higher errors. The symmetry lines over which the data is plotted are shown with the schematic triangle.}
\label{fig: moment_gf2}
\end{figure}

For two dimensions, we only have data from the VCA and from the QMC.  We have analyzed all of the moments in a similar fashion to what was done for the one-dimensional case, and we summarize our results in a series of figures.  We choose parameters corresponding to the second Mott lobe with $n=2$, and with a hopping that lies inside the lobe but close to the tip ($t/U=0.03$ and $\mu=1.45$). In this case, we might expect to see larger deviations from the sum rules. The quantum Monte Carlo results are run at a low temperature $\beta=80$.  In Figure~\ref{fig: moment_gf2}, we plot the exact results for the moment sum rules (as evaluated from the strong-coupling perturbation theory described above) versus the VCA and quantum Monte Carlo results from a maximum entropy analytic continuation. Since the momenta are now distributed through the two-dimensional Brillouin zone, we show plots for momenta along the high-symmetry lines of the triangle that runs from the origin at the $\Gamma$ point $(\Gamma=0)$ to the $M$ point along the diagonal direction $[M=(\pi,\pi)]$, to the $X$ point along the axial direction $[X=(\pi,0)]$. One can see for the low moments, the VCA approximation continues to work extremely well (once again, the VCA is evaluated as sums over the delta functions, and uses no broadening, which is why the moments are so accurate). But even in this case, we do start to see deviations for higher moments, and they are larger than they were for the one-dimensional case; this is expected due to the proximity to the tip of the Mott lobe at $t=0.035$~\cite{2dlobe}.  The QMC results, on the other hand, do show the right trends, but have a much larger quantitative error to the moments.

\begin{figure}[htb]
\vskip -1.0in
\begin{minipage}[h]{0.99\linewidth}
\centerline{\includegraphics[scale=0.34]{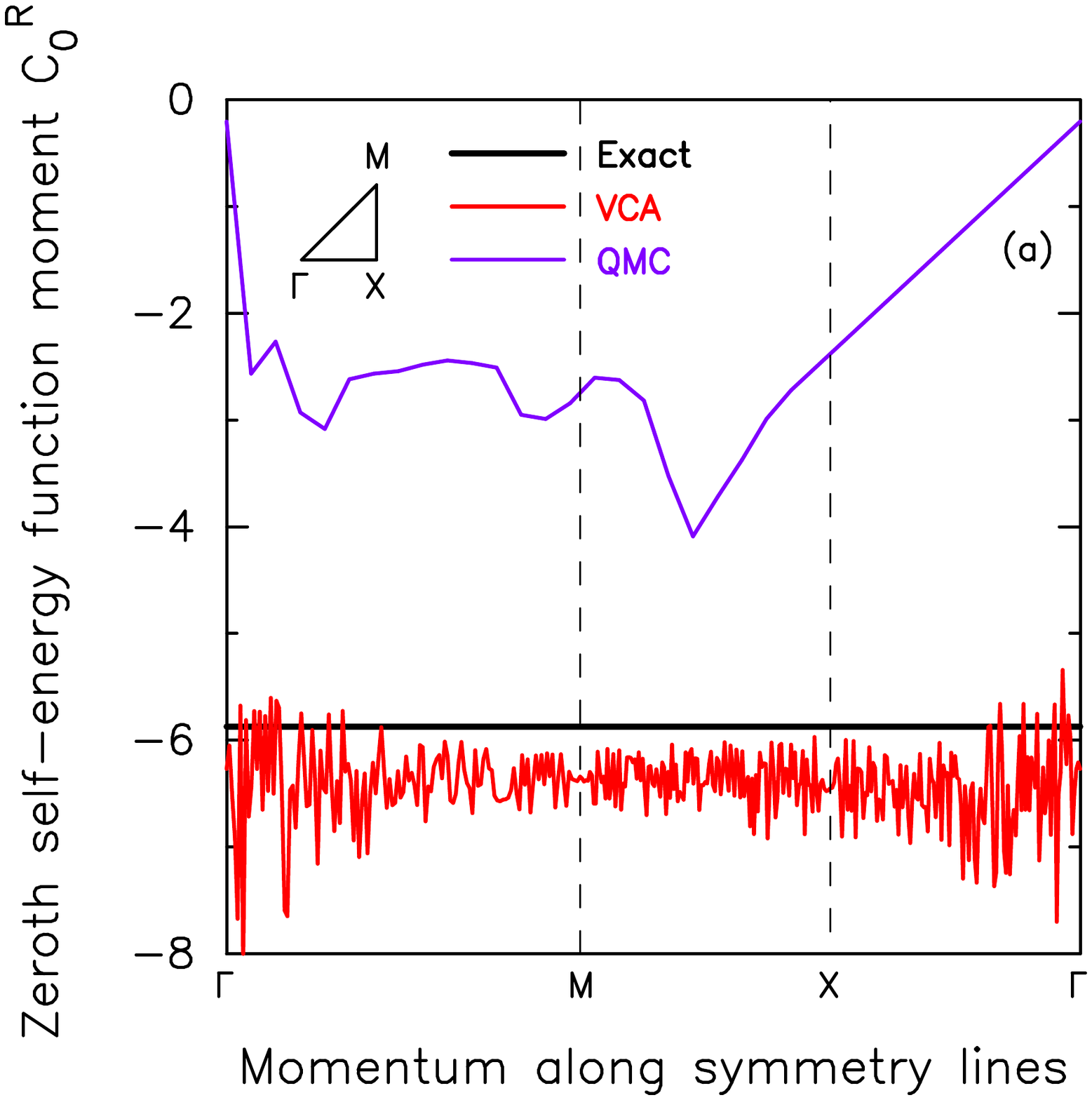}
\includegraphics[scale=0.34]{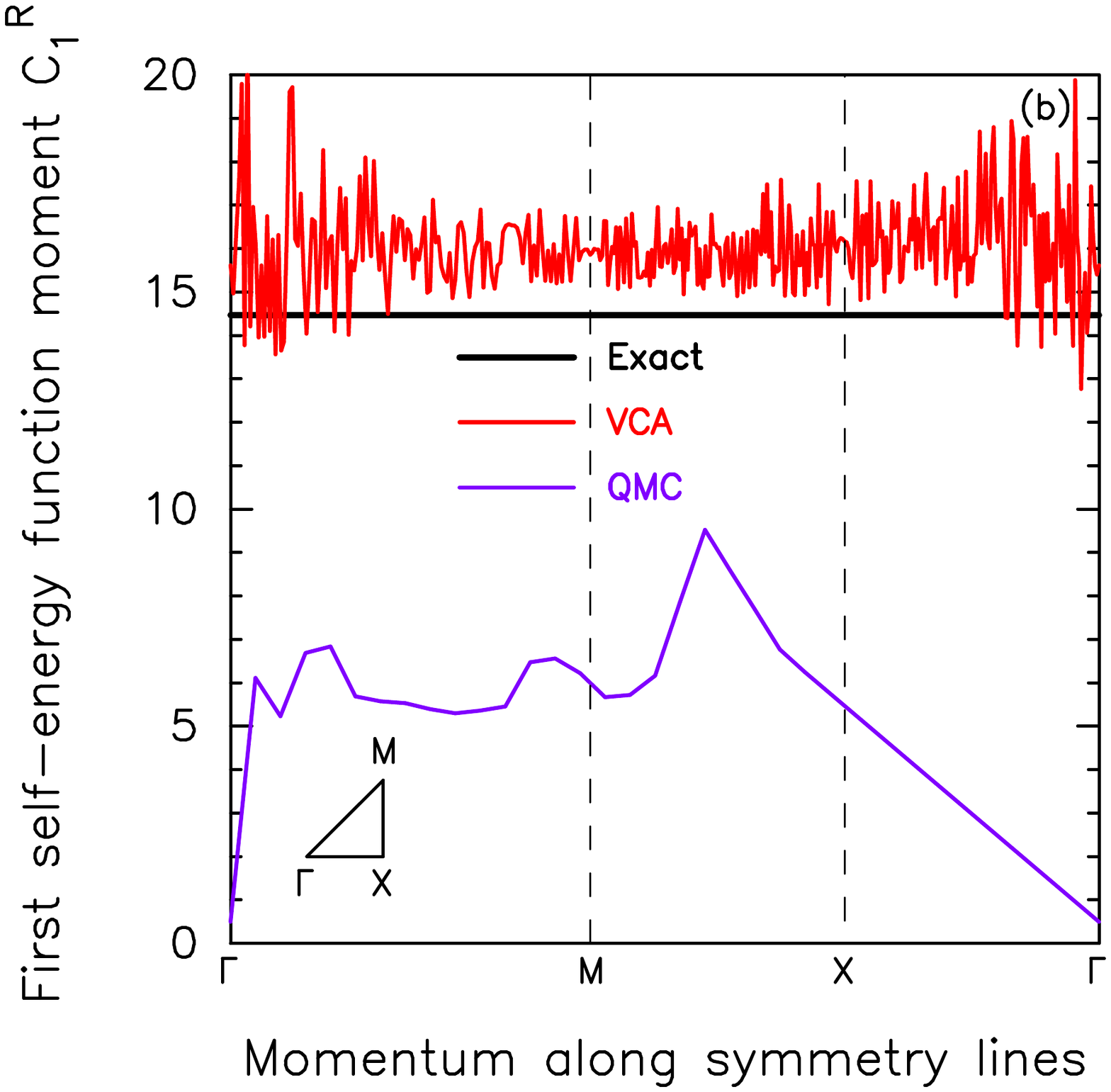}}
\end{minipage}
\caption{(Color online.) Zeroth (a) and first (b)  spectral moments of the retarded self-energy as a function of momentum for the two-dimensional Bose Hubbard model on a square lattice in the Mott-insulating phase
with $n=2$.  The parameters are $t/U=0.03$ and $\mu/U=1.45$, with energies measured in units of $U$.  We compare the exact result (black) to the variational cluster approach  (red line), and to the quantum Monte Carlo plus maximum entropy approach (purple). Deviations are visible for all approximations indicating the self-energy is not determined as accurately here.}
\label{fig: moment_se2}
\end{figure}

We continue with plots of the zeroth and first moments of the self-energy in Fig.~\ref{fig: moment_se2}. Here, we must go through the inversion procedure described above to extract the self-energy from the data for the Green's function. Hence, for the VCA, we must broaden the delta functions to create a smooth functional form for the Green's function.  We do this with a narrow broadening parameter to try to preserve the accuracy.  This gives rise to the amplitude of the oscillations in the moment data due to oscillations in the Green's function and then in the self-energy.  Once again, we do see the right trends in the data, but here the accuracy is fairly poor, especially for the QMC data.  The VCA data tends to have the right average behavior, but it's quantitative average is incorrect.  By looking at the traces of the self-energy itself, there are likely two reasons for the discrepancy.  The first, is that the frequency cutoff might be too low, and the second is that the self-energy seems to have regions of frequency where it  has strong frequency dependence, and this might not be properly captured by the approximations.

\section{Conclusions}

In this work, we have derived exact formulas for the first few spectral moments of the Bose-Hubbard model through third order for the retarded Green's function and through first order for the retarded self-energy. The results we derive are quite general, holding for inhomogeneous systems and for systems that have time dependence to the parameters in the Hamiltonian.  Sum rules can be quite useful in benchmarking different approximations, because their results are exact.  One challenge with the work here, is that for the bosonic case, the moments depend on correlation functions that must be determined for the interacting system, unlike in the fermionic version, where many of the correlation functions become trivial to evaluate. But, in the limit of strong coupling, for the Mott phase, these moments can be systematically found in a strong-coupling expansion, which appears to be quite accurate when compared to QMC results.  We concluded this work with numerical calculations for translationally invariant systems in equilibrium, where we could benchmark the accuracy of different numerical results.  Because we did this in the strong-coupling region, it comes as no surprise that the VCA turned out to be the most accurate approach, indicating that it is faithfully producing the moments of the spectral functions.  It is much more difficult for us to determine the point-wise accuracy of the spectral functions, though.  We did see that the numerical calculations appear to work best in one-dimension, where the moment sum rules are most accurate. 

We hope that this work will be used by others for benchmarking purposes of numerics and possibly for understanding qualitative changes in spectra as parameters change due to the constraints given by the sum rules.  As nonequilibrium techniques are developed for interacting bosonic systems, it will also be interesting to use these results for benchmarking of those calculations, since exact results for nonequilibrium systems are quite rare.

\acknowledgments

We acknowledge Peter Pippan for providing us with quantum Monte Carlo data and Satoshi Ejima for the density matrix renormalization group data. We acknowledge useful discussions with Enrico Arrigoni and Wolfgang von der Linden.  J.K.F. was supported by the National Science Foundation under grant number DMR-1006605. V.T. acknowledges the Department of Energy under grants numbered  DE-FG02-07ER15842
and DE-FG02-07ER46354. H.R.K. acknowledges the Indian DST. M.K. acknowledges support from the Austrian Science Fund under project number P24081-N16 and from the Austrian Marshall Plan Foundation. The collaboration between the US and India (J.K.F. and H.R.K.) was supported by the Indo-US Science and Technology Forum under a joint research center numbered JC-18-2009 (Ultracold atoms).

\end{document}